%% file: main.tex
\documentclass[journal]{IEEEtran}

\ifCLASSINFOpdf
\else
\fi
\usepackage{graphics} % for pdf, bitmapped graphics files
\usepackage{epsfig} % for postscript graphics files
\usepackage{mathptmx} % assumes new font selection scheme installed
\usepackage{times} % assumes new font selection scheme installed
\usepackage{amssymb}  % assumes amsmath package installed

\usepackage{amsmath}
\usepackage[linesnumbered,ruled]{algorithm2e}
\usepackage{pdfpages}
\usepackage{lipsum}

\usepackage{tabularx,booktabs}
\DeclareMathAlphabet{\mathcal}{OMS}{cmsy}{m}{n}

\usepackage{mathtools, nccmath}

\usepackage[nottoc]{tocbibind}
\usepackage{tocloft}
\usepackage{titletoc}
\usepackage{times} %times new roman 
\usepackage{mathptmx}
\usepackage{graphicx}
\usepackage{physics}

\usepackage{array}

\usepackage{multirow}
\usepackage{float} %for pictures floating

\DeclareMathAlphabet{\mathcal}{OMS}{cmsy}{m}{n}

\usepackage[noadjust]{cite} % for fixing multiple citations

\newcommand{\RN}[1]{%
  \textup{\uppercase\expandafter{\romannumeral#1}}%
}

\usepackage{mathtools}

\usepackage{caption}
\usepackage{subcaption}

\begin{document}
\title{Real-Time Forward Kinematics and Jacobians for Control of an MRI-Guided Magnetically Actuated Robotic Catheter}

\author{Ran Hao, Yuttana Itsarachaiyot, %~\IEEEmembership{Student Member,~IEEE,} 
         Yen-Chun Chen, %~\IEEEmembership{Fellow,~OSA,}
        and M. Cenk \c{C}avu\c{s}o\u{g}lu%~\IEEEmembership{Life~Fellow,~IEEE}% <-this % stops a space
\thanks{This work was supported in part by the National Science Foundation under grants CISE IIS-1563805, and ENG IIP-1700839, and National Institutes of Health under grants R01 HL153034, and R01 HL163991.}% <-this % stops a space
\thanks{R. Hao, Y. Itsarachaiyot, Y.-C. Chen, and M. Cenk \c{C}avu\c{s}o\u{g}lu are with the Department of Electrical, Computer, and Systems Engineering, Case Western Reserve University, Cleveland, OH, 44106. They can be reached via email at rxh349@case.edu, yxi41@case.edu, yxc1873@case.edu and mcc14@case.edu respectively. This paper was presented in part at the 2023 IEEE/RSJ International Conference on Intelligent Robots and Systems (IROS), Detroit, USA. }% <-this % stops a space
% \thanks{Manuscript received April 19, 2005; revised August 26, 2015.}
}

% \markboth{Journal of \LaTeX\ Class Files,~Vol.~14, No.~8, August~2015}%
% {Shell \MakeLowercase{\textit{et al.}}: Bare Demo of IEEEtran.cls for IEEE Journals}

\maketitle

\begin{abstract}

This paper presents a forward kinematics and analytical Jacobian computation approach for real-time control of a novel magnetic resonance imaging (MRI)-actuated robotic catheter. The MRI-actuated robotic catheter is modeled as a series of rigid and flexible segments and actuated by magnetic torques generated on a set of current-carrying microcoils embedded on the catheter body by the magnetic field of the MRI scanner.  First, a real-time forward kinematic modeling approach of the robotic catheter employing the static Cosserat-rod theory is presented. Second, the analytical calculation approach of the forward kinematic Jacobians of the proposed forward kinematic model is presented. The accuracy, reproducibility, and computational efficiency of the proposed methods are evaluated using a robotic catheter prototype with a single coil set, where catheter tip trajectories collected by a catadioptric stereo camera tracking system are validated using the desired tip trajectories. Experimental results demonstrate that the proposed method can successfully control the catheter in an open loop to perform complex trajectories with real-time computational efficiency, paving the way for accurate closed-loop control with real-time MR-imaging feedback.

\end{abstract}


\IEEEpeerreviewmaketitle

\section{Introduction}\label{section:introduction}
% In this paper, a real-time free-space kinematic modeling approach for a MRI-actuated robotic catheter is proposed.

Robotic intravascular cardiac catheter ablation systems have been proposed to improve the ablation outcomes by offering improved accuracy, efficiency and efficacy in treating atrial fibrillation, compared to traditional catheter ablation procedures\cite{kanagaratnam2008experience,kesner2011position,saliba2008atrial,kesner2014robotic}. In \cite{liu2016modeling,liu2017design,greigarn2015pseudo}, a novel magnetic resonance imaging (MRI)-actuated robotic intravascular catheter system was proposed, aiming to allow the atrial fibrillation ablation procedure to be performed under real-time intraoperative MRI guidance. The MRI-actuated robotic catheter is embedded with multiple sets of electromagnetic micro-coils. The bending of the robotic catheter is controlled by the Lorentz forces generated by the static magnetic field of the MRI scanner, eliminating the actuation friction and backlash that exist in methods where the actuators are placed outside the patient’s body (such as tendon-driven and hydraulic actuation), and resulting in high actuation bandwidth \cite{Tuna18}. A finite-differences-based kinematic model has been proposed for controlling the MRI-actuated robotic catheter\cite{liu2016modeling,liu2017iterative}. While it has demonstrated high accuracy, it is not computationally efficient for real-time catheter control applications.

This paper focuses on derivation of an analytical Jacobian of the kinematics to facilitate the real-time kinematic modeling of the MRI-actuated robotic catheter using Cosserat-rod theory and its application in real-time open-loop control. The main contribution of this paper includes a virtual angular velocity based approach for calculating the set of forward kinematics Jacobians relating the changes of the control inputs, i.e., the coil actuation currents and catheter insertion, to the changes of the catheter tip position, orientation, and curvatures using the Cosserat-rod forward kinematic model. Specifically, the proposed derivation is based on the direct integration of the rotation matrix without losing the structure of the 3D rotation group $SO(3)$. The real-time computational efficiency of the presented approach is demonstrated in experiments.

In this paper, the iterative-Jacobian-based inverse kine- matic method is employed to control the deflection of the MRI-actuated robotic catheter. The iterative-Jacobian-based inverse kinematics method employs the derived analytical Jacobians and iteratively updates the actuation inputs given the desired catheter tip configuration displacements. A single- coil set robotic catheter prototype is used to experimentally evaluate the performance of the proposed method, where the 3D catheter tip position trajectories collected under open-loop control are compared with the desired trajectories. A catadioptric stereo camera system is used to track the catheter deflections in the experiments. Experimental results show that the proposed approach can accurately control the robotic catheter with real-time computational efficiency.

The rest of this paper is organized as follows. Related studies in the literature are presented in Section \ref{section:relatedstudies}. The forward kinematic modeling of the MRI-actuated robotic catheter based on Cosserat-rod theory is given in Section \ref{section:cosseratmodel}. The analytical derivations of the forward kinematic Jacobians are presented in Section \ref{sec:parameterderivatives}. Experimental validation of the proposed methods are provided in Section \ref{section:experiment}, followed by the discussions and conclusions in Section \ref{section:discussion} and \ref{section:conclusion}, respectively.

\section{Related Studies}\label{section:relatedstudies}

Cosserat rod theory is widely used in kinematic modeling of continuum robots \cite{rucker2010geometrically, rucker2011statics, edelmann2017magnetic, antman2005, trivedi2008geometrically, dupont2009design}. The Cosserat rod model is a mathematical model used to describe the behavior of slender elastic rods. Unlike traditional rod models, the Cosserat rod model takes into account both the bending and twisting deformations of the rod. Specifically, the model considers the rod as a continuous distribution of material points, each having its own position, orientation, and twist angle.
% The rod is represented as a set of interconnected line segments, where each segment has its own orientation and twist. 
The deformation of the rod is described by a set of partial differential equations that relate the forces, moments, and torques acting on the rod to its curvature, torsion, and twist.

Kinematic modeling and control of continuum robots has been widely investigated in the literature \cite{rucker2010geometrically, rucker2011statics, edelmann2017magnetic, mahvash2010stiffness, tummers2023cosserat, till2015efficient, till2017elastic, grazioso2019geometrically, alqumsan2019robust, orekhov2017modeling, barrientos2021real}. Mahvash and Dupont \cite{mahvash2010stiffness} propose a special Cosserat rod model for computing deformation due to external loading, enabling computationally efficient calculation of robot deflection. Stiffness control is implemented using an iterative method to solve for actuator positions that achieve the desired tip force. In \cite{tummers2023cosserat}, Tummers \textit{et al.} focus on the modeling of continuum robots, specifically Tendon-Actuated Continuum Robots (TACRs), from both Newtonian and Lagrangian perspectives, in which the Lagrangian approach involves a systematic reduction process of the Cosserat rod model. 
Till \textit{et al.} \cite{till2015efficient, till2017elastic} formulate a kinematic model treating parallel continuum robots as multiple Cosserat rods with coupled boundary conditions, enabling real-time interactive simulation, motion planning, design optimization, and control. 
Lilge and Burgner-Kahrs \cite{lilge2022kinetostatic} develop a Cosserat rod-based kinetostatic modeling framework for tendon-driven parallel continuum robots, enabling solutions for forward, inverse, and velocity kinetostatic problems.
Doroudchi and Berman \cite{doroudchi2021configuration} propose an inverse dynamic control approach using geometrically exact Cosserat rods for 3D configuration tracking.
Samei and Chhabra \cite{samei2023fast} create a geometric framework using the finite difference method for the dynamic Cosserat rods model for real-time applications.
Li \textit{et al.} \cite{li2023piecewise, li2023discrete} explores a closed-loop control architecture for soft manipulators using a reformulated Cosserat static model, solving highly nonlinear partial differential equations of Cosserat-based models.
Boyer \textit{et al.} \cite{boyer2022statics} explores the relationship between optimal control theory and Cosserat beam theory for solving forward and inverse dynamics of continuous manipulators.
Ghazbi \textit{et al.} \cite{ghazbi2020jacobian} introduces cooperative continuum robots, presenting kinematic modeling and Jacobian derivations that combine the DH approach with Cosserat rod theory for precise kinematic modeling.

In \cite{rucker2011computing}, Rucker and Webster propose a computation method for the forward kinematic Jacobians by propagating the initial value problem (IVP) partial derivatives along the robot body, where the IVP derivatives are calculated using the finite difference approach.  Liu et al.\cite{liu2017iterative} propose an iterative-Jacobian based open loop control method based on the finite- difference-based kinematic model of the MRI-actuated robotic catheter, where the closed-form calculations of the Jacobians are derived based on the finite element kinematic model. Edelmann et al. \cite{edelmann2017magnetic} propose a control system for closed- loop control of a magnetically actuated continuum robot with permanent magnets embedded on the robot body. The forward kinematics Jacobians are derived analytically based on the Cosserat kinematic model.

The objective of this paper is to investigate and evaluate the efficiency and effectiveness of analytical computation of the forward kinematic Jacobians and kinematic modeling for real-time open-loop control of the MRI-actuated robotic catheter. Specifically, a virtual angular velocity vector is proposed to preserve the structure of the rotation matrix during the forward integration along the catheter body for solving the initial value problem. The forward kinematic Jacobians are then derived analytically using the virtual angular velocity vector representation of the rotation matrix. In this work, the forward kinematic Jacobians are derived in analytically, similar to the analytical kinematic Jacobian calculations proposed in \cite{rucker2011computing}; however, unlike the method proposed in \cite{rucker2011computing}, the proposed kinematic Jacobian derivation provides explicit closed form definitions of both the IVP and BVP Jacobians based on the virtual angular velocity vector, which further facilities the derivation propagation without the need for any numerical calculations. In \cite{edelmann2017magnetic}, the forward kinematic Jacobians for the rigid and flexible segments are derived analytically using the quaternion representation of the rotation. In this paper, a rotation matrix based analytical derivation method for efficient calculations of the forward kinematic Jacobians are proposed. Specifically, a virtual angular velocity vector is proposed for the computation and integration of the rotation matrix along the catheter body without truncation and roundoff error, which is an alternative method of integrating rotation matrices compared to the method proposed in \cite{8392463}. In addition, we demonstrate that the iterative-Jacobian based open-loop control implemented using the analytically calculated Jacobians achieves better computational efficiency, compared to the finite-differences based modeling approach proposed in \cite{liu2017iterative}, \cite{rucker2011computing}, enabling a real-time control scenario. To the best of our knowledge, this is the first work that calculates the forward kinematic Jacobians directly in the rotation group $SO(3)$ for the Cosserat-rod forward kinematic models.

The computational efficiency and accuracy of the proposed forward kinematic Jacobian derivation approach are validated in the experiments. We demonstrate that the proposed method delivers real-time performance with comparable accuracy, compared to the results shown in \cite{liu2016modeling}, paving the way for the closed-loop control system with real-time MRI guidance information.

\input{crm_kinematics}

\input{experiment}

\section{Discussion}\label{section:discussion}

In Table~\ref{tb:table_accuracy_crm_pebax}, the reported average RMSEs of the CRM based open-loop control trajectories are in the range of 3.66 mm - 6.49 mm. There are several potential error sources, including the camera tracking errors and the offset and drift type of errors, which can be reduced by applying closed-loop control strategies. In Table~\ref{tb:table_reproduce_crm_pebax}, the mean of the RMSEs between the observed trajectories and the desired tip trajectories after eliminating the positional and orientation offsets are 1.81 mm, 2.29 mm, 3.35 mm, and 2.72 mm for circle, lemniscate, rectangle, and butterfly trajectory shapes, respectively. The variability between multiple trials of the observed trajectories and the random reference tip trajectories is below 1 mm with low variances, indicating good repeatability.

The offsets between the starting point of the observed trajectories in Fig.~\ref{fig:example_trajectory}, are due to plastic deformation of the catheter tubing material during the repeated bending of the catheter. This issue can be alleviated by using different types of tubing materials for the catheter body, which will be the subject of our future work.

As shown in Fig.~\ref{pic:comparison}, the mean of the RMSEs for the FDM-based trajectories are 2.43 mm, 4.09 mm, 4.78 mm, and 7.46 mm, for circle, lemniscate, rectangle and butterfly, respectively. The mean of the RMSEs for the CRM-based trajectories are 2.84 mm, 5.33 mm, 5.97 mm, and 6.44 mm, for circle, lemniscate, rectangle and butterfly, respectively. The accuracy of the CRM-based control method is comparable with the FDM based control method, where the difference between the RMS errors are below 2 mm for all trajectory shapes. The FDM based control method achieves slightly higher accuracy for circle, lemniscate, and rectangle trajectory shapes, whereas, the Cosserat rod based method achieved higher accuracy for the more complex butterfly trajectory. For closed-loop control of the MRI-actuated robotic catheter with real-time MRI feedback information, the run time per step of the open-loop control should be significantly smaller than the targeted image acquisition time of 16 ms or catheter localization \cite{franson2021system}. The proposed method, with is 4.1 ms run time per control step, is sufficiently fast for real-time closed-loop control performance.

\section{Conclusion}\label{section:conclusion}

In this paper, the real-time open-loop control of an MRI-actuated robotic catheter is studied. An analytical Jacobian is proposed for real-time computation of the Cosserat-rod based kinematic model of the MRI-actuated robotic catheter. The proposed Jacobian method and kinematic modeling approach are employed by the iterative-Jacobian-based inverse kinematic method for open-loop control of the deflection for the MRI-actuated robotic catheter.

Performance evaluation of the proposed kinematic model and Jacobian calculation is presented, where 4 distinct sets of trajectory shapes (circle, lemniscate, rectangle, and butterfly) are used to evaluate the accuracy, reproducibility, and computational efficiency of the proposed methods. Experiment results show that the proposed method provides comparable accuracy and repeatability of trajectories while improving the computational efficiency, compared to the numerical Jacobian and finite-element based kinematic modeling methods. The proposed Jacobian and kinematic modeling approach is efficient and accurate, enabling a real-time open-loop control scheme, which is a crucial first step for close-loop control implementation with real-time MR-imaging feedback. Our future work will focus on the implementation and validation of close-loop control of the MRI-actuated robotic catheter.

% \appendices

% % use section* for acknowledgment
% \section*{Acknowledgment}

\ifCLASSOPTIONcaptionsoff
  \newpage
\fi



%\bibliographystyle{IEEEtran}


% \end{thebibliography}

\ifCLASSOPTIONcaptionsoff
  \newpage
\fi

% \appendices
% \input{Appendix}

\bibliographystyle{IEEEtran}
\bibliography{references}
\vskip 0pt plus -1fil


%\vfill

% \enlargethispage{-5in}

\end{document}

%% file: crm_kinematics.tex
\section{Kinematic Model of The MRI-Actuated Robotic Catheter}\label{section:cosseratmodel} 

\subsection{Catheter Prototype Description}\label{subsection:catheter_prototype}

\begin{figure}[t]
\includegraphics[scale=0.2]{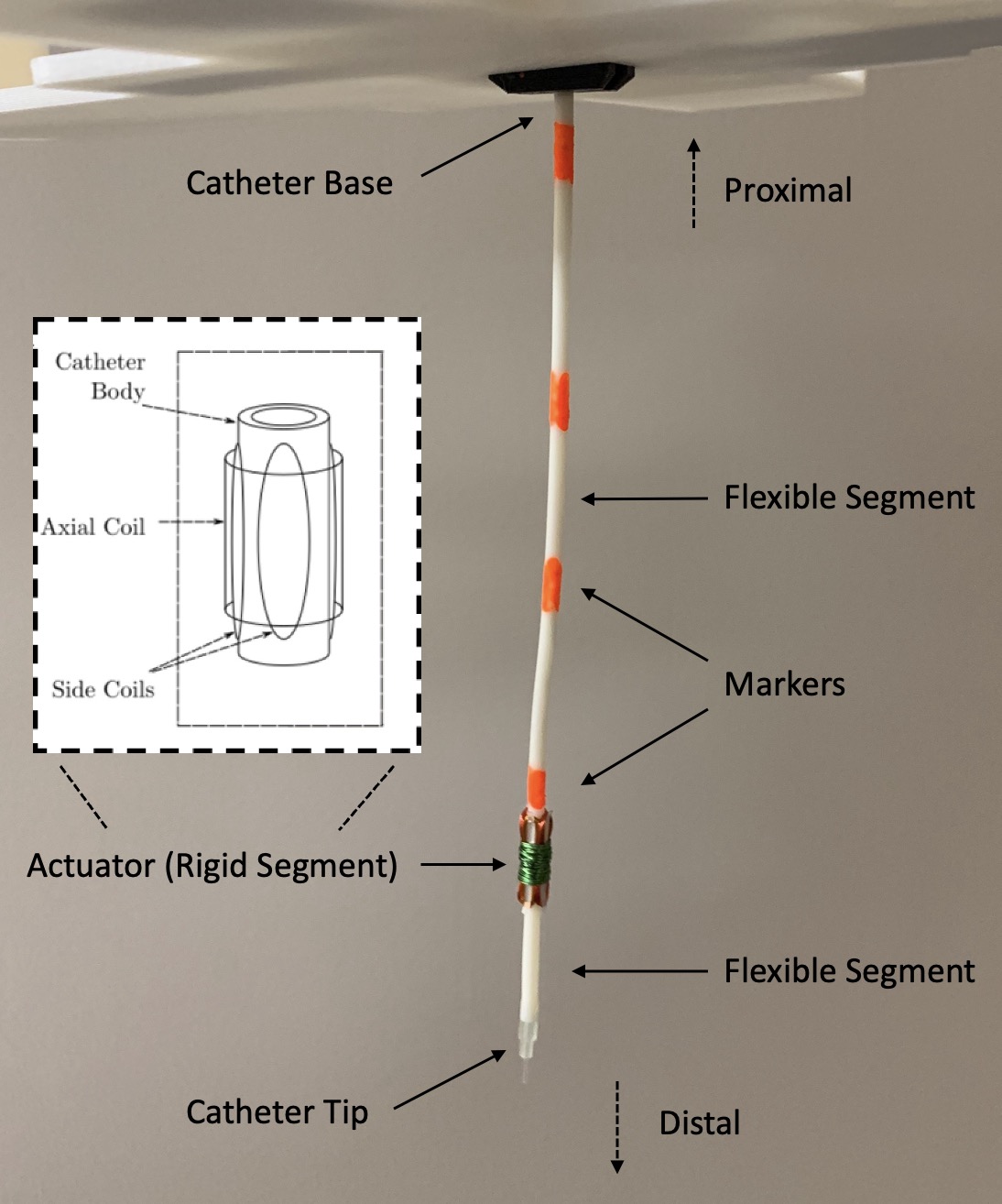}
\centering
\caption{The prototype of one coil set MRI-actuated robotic catheter. The actuator contains 1 axial coil and 2 orthogonal side coils. The design and parameter optimization of the robotic catheter can be found in \cite{greigarn2015pseudo,liu2016modeling,liu2017design}.}
\label{pic:catheter_prototype}
\end{figure}

The MRI-actuated robotic catheter \cite{liu2016modeling,liu2017design, greigarn2015pseudo}, as shown in Fig.~\ref{pic:catheter_prototype}, is embedded with one current carrying coil actuator. The actuator contains one axial coil, and two orthogonal side coils \cite{liu2016modeling}. The catheter body is made of polymer tubing (Pebax 35D, Arkema Inc, PA, USA). The catheter tubing has the inner radius of 0.8 mm and outer radius of 1.3 mm. The total length of the catheter is 146.0 mm while the length of the coil set is 13.0 mm.

The orange markers on the catheter as shown in Fig.~\ref{pic:catheter_prototype} are used to estimate an initial catheter shape using an external camera and catadioptric stereo tracking system \cite{Russell_icra2016}. The initial shape is used to estimate the rest curvature ($u^*$) of the catheter body for the Cosserat-rod model presented in Section \ref{subsection:catheter_rod_modeling}.

\subsection{Cosserat Rod Modeling of the MRI-Actuated Robotic Catheter}\label{subsection:catheter_rod_modeling}

\begin{figure}[tb]
\includegraphics[scale=0.15]{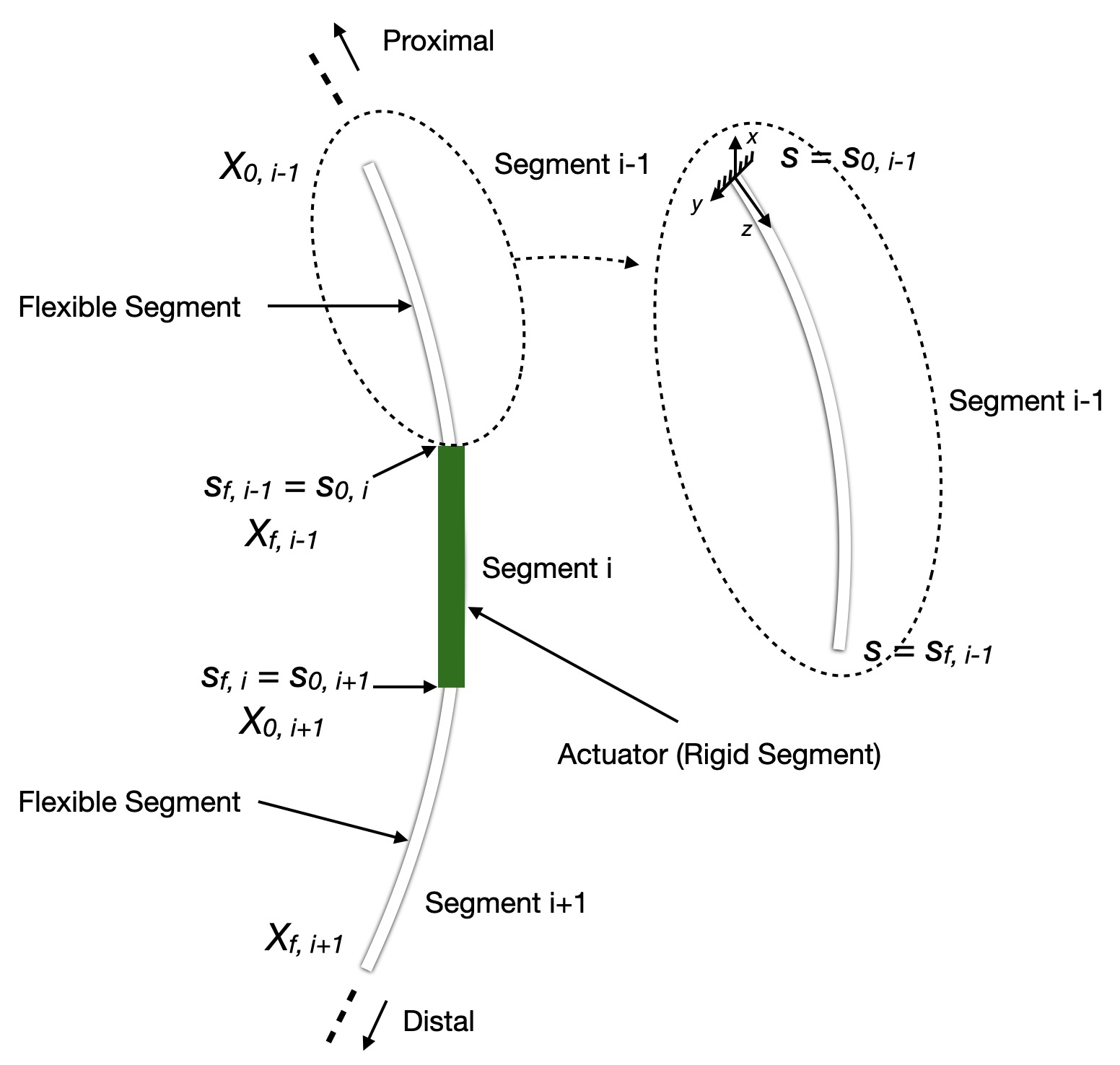}
\centering
\caption{An example segment ordering of the MRI-actuated robotic catheter with one rigid segment $i$ connected by two flexible segments $i-1$ and $i+1$. $s$ denotes the length parameter along the catheter body. $X$ denotes the catheter state variables, representing the position $p(s)\in \mathbb{R}^3$, orientation $R(s) \in SO(3)$, and curvature $u(s) \in \mathbb{R}^3$ along the length of the catheter, parameterized by the length parameter $s$. }
\label{pic:segment_order}
\vspace{-0.2in}
\end{figure}

The MRI-actuated robotic catheter (Fig.~\ref{pic:catheter_prototype}) is steered by the magnetic moments generated on a set of current carrying actuation coils mounted on the catheter body by the static magnetic field of an MRI scanner \cite{liu2017design}.  Each actuator on the catheter contains one axial and two orthogonal side coils, as given in Fig.~\ref{pic:catheter_prototype}. The actuation inputs of the robotic catheter is then given as $z = [z^c, z^l]^T$, where $z^c$ is the vector of actuation currents, and $z^l$ is the inserted length of the catheter. Let $z^c_i \in \mathbb{R}^3$ denote the actuation currents of the $i$-th actuator, the magnetic moment $\tau_i$ generated by the $i$-th actuator is then given as 
\begin{equation}
\tau_i = (N_i A_i z_{i}^c)^\wedge\, B_i = -{B_i}^\wedge\, (N_i A_i) z_{i}^c, 
\label{eq:tau_def}
\end{equation}
where the $^\wedge$ operator maps a vector in $\mathbb{R}^3$ to so(3). $N_i$ and $A_i$ are 3 by 3 diagonal matrices whose diagonal elements are the number of winding turns and the cross sectional areas, respectively. The static magnetic field $B_s$ of the MRI scanner expressed in the body frame of the $i$-th actuator is given as $B_i = R_{i}^T B_s$, where $R_{i}$ denotes the rotation matrix of the frame attached to the $i$-th actuator relative to the spatial frame. 

In this study, the kinematics of the robotic catheter is modeled using the Cosserat rod theory as a series of flexible and rigid segments. A schematic of the segment ordering is shown in Fig. \ref{pic:segment_order}.
where $X$ denotes the catheter state variables, representing the position $p(s)\in \mathbb{R}^3$, orientation $R(s) \in SO(3)$, and curvature $u(s) \in \mathbb{R}^3$ along the length of the catheter, as parameterized by the length parameter $s$. 
For the $i$-th flexible segment,  $X_{0,i-1}$ and $X_{f,i-1}$ denote the starting point and the end point of the segment, respectively.
In the Cosserat rod model (CRM), calculation of the equilibrium configuration of the catheter under external loads, including actuation moments, $\tau_i$, and the contact force at the tip of the catheter, $f_{tip}$, involves the solution of a boundary value problem (BVP), over the inserted length of the catheter $s \in [0,z^l]$.

Specifically, for a given flexible segment (e.g., segment $i-1$ in Fig. \ref{pic:segment_order}), the catheter state variables satisfy the CRM differential equations (as given, e.g., by Antman \cite{antman2005}, Rucker et al. \cite{rucker2010geometrically}, or Edelmann et al. \cite{edelmann2017magnetic}):
\begin{equation}
\begin{split}
  \dot{p}(s) &= R(s)e_3, \\ \dot{R}(s) &= R(s)\widehat{u}(s), \\
  \dot{u}(s) &= \dot{u}^*(s) - K^{-1}(s) \Bigl( \bigl(\widehat{u}(s)K(s) + \dot{K}(s)\bigl)\bigl(u(s)-u^*(s) \mathrlap{\bigl)} \\
  & + \widehat{e}_3R^T(s) f_{cum}(s) + R^T(s)l  \Bigl),
\end{split}
\label{eq:p_R_u_dot}
\end{equation}
%\begin{equation}
%\begin{split}
%   \dot{p}_{s, i-1} &= R_{s, i-1}e_3; \ \ \dot{R}_{s, i-1} = R_{s, i-1}\widehat{u}_{s, i-1}; \\
%   \dot{u}_{s, i-1} &= \dot{u}^*_{s, i-1} - K^{-1}_{s, i-1} \Bigl( \bigl(\widehat{u}_{s, i-1}K_{s, i-1} + \dot{K}_{s, i-1}\bigl)\bigl(u_{s, i-1}-u^*_{s, i-1}\bigl) \\
%   & + \widehat{e}_3R^T_{s, i-1} f_{cum} + R^T_{s, i-1}l  \Bigl),
%\end{split}
%\label{eq:p_R_u_dot}
%\end{equation}
subject to the boundary conditions
\begin{equation}
\begin{split}
p(0) &= p_0; \ \ R(0) = R_0; \\
\tau_{res} &= K\left( u(s=z^l) - u^*(s=z^l) \right) - l_{tip}= \mathbf{0}.
\end{split}
\label{eq:moment_boundary}
\end{equation}
$p_0$ is the location of the base of the catheter (i.e., the entry point) and $R_0$ is the orientation of the catheter at the base. $e_3 = [0,0,1]^T$ is the unit vector in $z$ direction. $\tau_{res}$ is the residual moment calculated at the catheter tip. $f_{cum}(s) = \int_s^{z^l} f(\sigma) d\sigma  + f_{tip}$ is the accumulated external force applied on the catheter body from $s$ to the distal tip of the catheter. $l_{tip}$ is the moment applied at the tip of the catheter; $l_{tip} = \tau_i$ if the $i$-th actuator is the last segment of the catheter, and $l_{tip} = 0$ if the last segment of the catheter is a flexible segment.
% $$K(s) =
% \begin{bmatrix}
%     E(s)I(s)    & 0         & 0 \\
%     0           & E(s)I(s)  & 0 \\
%     0           & 0         & G(s)J(s)     
% \end{bmatrix}$$
$u^*$ is the initial curvature of the flexible segment at rest. $K(s) = diag([E(s)I(s), E(s)I(s) , G(s)J(s)])$ is the stiffness coefficient matrix. $E(s)$ and $G(s)$ are the Young's modulus and shear modulus, respectively. $I(s)$ is the second moment of area and $J(s)$ is the polar moment of inertia of the catheter cross section.
For a rigid segment (e.g., rigid segment $i$ in Fig. \ref{pic:segment_order}), the catheter state variables are passed through the segment via the following equations
\begin{equation}
\begin{split}
p(s_{0, i+1}) &= p(s_{f, i-1}) + L_i R(s_{f, i-1}) e_3, \\  
R(s_{0, i+1}) &= R(s_{f, i-1}), \\
u(s_{0, i+1}) &= u^*(s_{0, i+1}) + \\ & K_{i+1}^{-1} \Bigl( K_{i-1}\left(u(s_{f, i-1}) - u^*(s_{f, i-1}) \right) - {\tau_i}  \mathrlap{\Bigl)}\ \ ,\\ 
\end{split}
\label{eq:rigid_PRu}
\end{equation}
where $L_i$ denotes the length of the rigid segment, and $e_3 = [0, 0, 1]^T$. 

% Cosserat rod model (CRM) is one of the nonparametric methods representing the deformation of a long object. All internal stresses and the pose and orientation of every station point along the length of the object are taken into account. 
% The derivations of the model are based on Antman's work \cite{antman2005} and Rucker et al. \cite{rucker2010geometrically} provide a statement of a set of differential equations derived from a moment balance and the strain energy describing how the position and orientation change along the object's length. 

\begin{figure}[tb]
\includegraphics[scale=0.225]{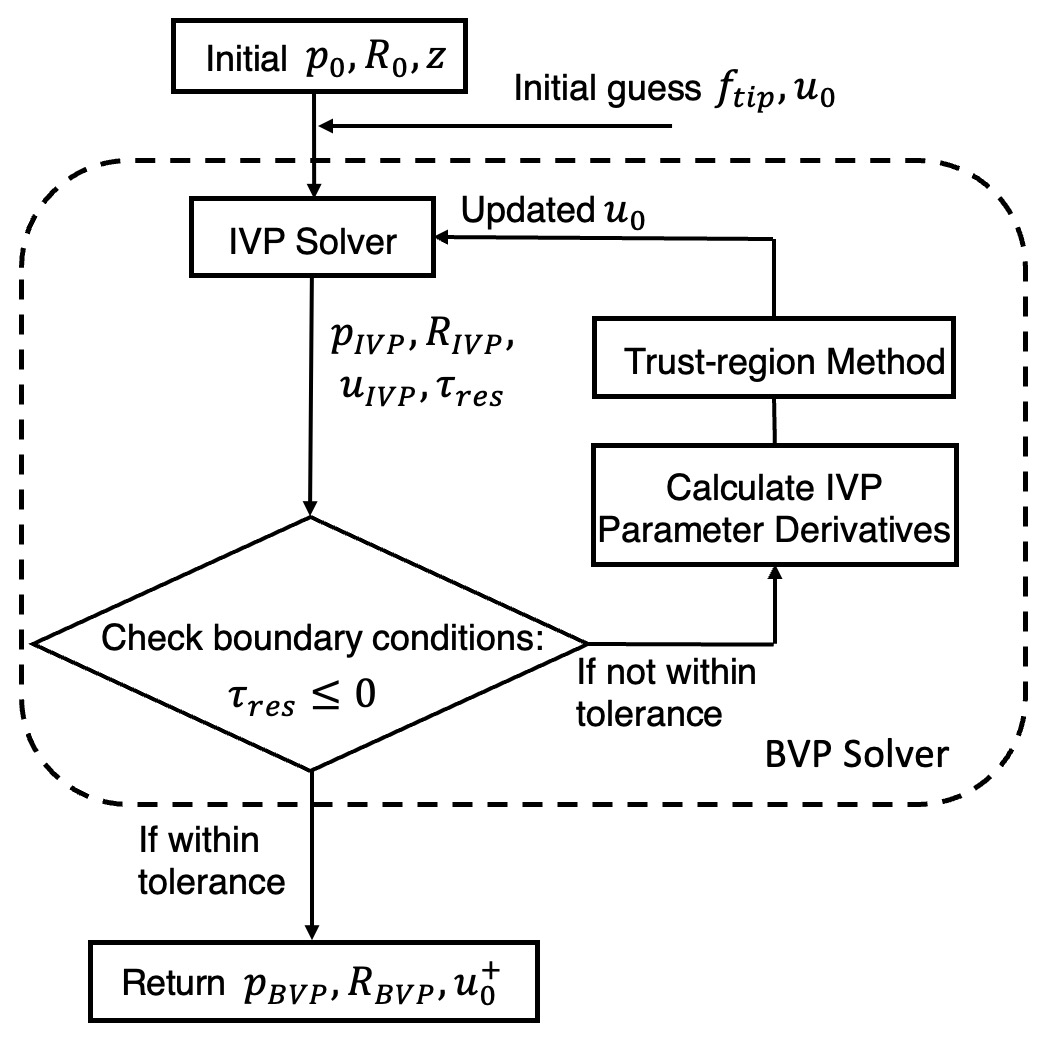}
\centering
\caption{The diagram of the BVP solver for the Cosserat-rod based kinematic model of the MRI-actuated robotic catheter.}
\label{pic:BVPdiagram}
\vspace{-0.2in}
\end{figure}

The BVP specified by (\ref{eq:p_R_u_dot}-\ref{eq:rigid_PRu}) is typically solved by using the Shooting Method (inner loop in Fig. \ref{pic:BVPdiagram}).  Specifically, in the Shooting Method, the BVP is iteratively solved by solving the initial value problem (IVP) given by the differential equation (\ref{eq:p_R_u_dot}) with initial conditions 
\begin{equation}
p(0) = p_0; \ \ R(0) = R_0; \ \ u(0) = u_0,
\label{eq:IVP_IC}
\end{equation}
where a nonlinear equation solving algorithm is employed to find the unknown initial curvature $u_0$ at the base of the catheter that satisfies the known boundary conditions (\ref{eq:moment_boundary}) at the distal tip of the catheter. 
% In this study, Trust-Region-Dogleg algorithm \cite{powell1970} is employed as the nonlinear equation solver.

The catheter state variables at the distal tip of the catheter calculated by the IVP solver through the integration of (\ref{eq:p_R_u_dot}) over the length of the catheter will be denoted as
\begin{equation}
[p_{\mathrm{IVP}},\ R_{\mathrm{IVP}},\ u_{\mathrm{IVP}}] =
\mathrm{IVP}(p_0,\ R_0,\ u_0,\ f_{tip},\ z),
\label{eq:IVP_equation}
\end{equation}
for given initial position $p_0$, initial rotation matrix $R_0$, initial curvature $u_0$, tip force $f_{tip}$ and actuation inputs $z$.
$p_{\mathrm{IVP}}$, $R_{\mathrm{IVP}}$ and $u_{\mathrm{IVP}}$ denote the position, rotation matrix, and curvatures computed by the IVP solver at the tip of the catheter.

The BVP function can then be expressed as
\begin{equation}
[p_{\mathrm{BVP}},\ R_{\mathrm{BVP}},\ u^+_0] =
\mathrm{BVP}(p_0,\ R_0,\ f_{tip},\ z),
\label{eq:BVP_function}
\end{equation}
where $p_{\mathrm{BVP}}$, $R_{\mathrm{BVP}}$ denote the catheter tip position and rotation matrix, respectively, and $u^+_0$ denotes the initial curvature, computed by the nonlinear equation solver as part of the BVP solution when the boundary conditions (\ref{eq:moment_boundary}) are satisfied.

As shown in Fig. \ref{pic:BVPdiagram}, the BVP solver calculates the tip position $p_{\mathrm{BVP}}$ and rotation matrix $R_{\mathrm{BVP}}$ by solving the IVP of the given initial position $p_0$, initial rotation matrix $R_0$, actuation inputs $z$, and the initial guesses of the initial curvature $u_0$. In this study, the BVP uses Trust-Region-Dogleg method \cite{powell1970} as the nonlinear equation solver for the root finding process.

\section{Forward Kinematic Jacobians}
\label{sec:parameterderivatives}
In this section, we present an analytical method for the calculation of the forward kinematic Jacobians defined in the boundary value problem in (\ref{eq:BVP_function}). Specifically, let $\phi \in \mathbb{R}^{n_{\phi}}$ denote the parameter vector which includes the parameters of actuation inputs $z$, initial curvature $u_0$, initial position $p_0$, and initial orientation $R_0$.
Let $X$ denote the catheter state variables representing the position, orientation, and curvature along the length of the catheter, as given in Fig. \ref{pic:segment_order}. 
We first define the initial value problem (IVP) parameter derivatives 
\begin{equation}
\centering
    J_{\mathrm{IVP}} = \partial X_{IVP} / \partial \phi 
    \label{eq:JIVP}
\end{equation}
as the derivatives of the catheter state variables at the catheter tip $X_{\mathrm{IVP}}$ w.r.t. the parameters $\phi$. The proposed method derives $J_{\mathrm{IVP}}$ analytically by integrating $ \partial X_{IVP} / \partial \phi$ along the catheter flexible and rigid segments to the tip of the robotic catheter at a given catheter configuration.

% The derivatives of the tip position w.r.t. the parameters $\partial p_{\mathrm{IVP}} / \partial \phi $ and the derivatives of the tip curvature w.r.t. the parameters $\partial u_{\mathrm{IVP}} / \partial \phi$ are derived analytically by integrating throughout the catheter body.
% We can define the forward kinematics parameter derivatives as $J_X =[\partial  p_{\mathrm{IVP}}/ \partial \phi, \partial u_{\mathrm{IVP}} / \partial \phi]$. The proposed method integrates the parameter derivatives $J_X$ throughout the catheter flexible and rigid segments to the tip of the robotic catheter at a given catheter configuration.
% The derivatives of the rotation w.r.t. the parameters $\partial R / \partial \phi$ is integrated along the catheter body for the calculation of the positional and curvature derivatives. 
\subsection{Virtual Angular Velocity Vector}
The direct integration of the derivatives of the rotation $R$ w.r.t. the parameters cannot guarantee the structure of SO(3) for $R$. In this paper, we define a virtual angular velocity vector $w_{\phi_j} \in \mathbb{R}^{3}$ for each element of $\phi_j$, where $\phi_j$ indicates the $j$-th value of the parameter vector $\phi$, $j = 1, 2, ...,n_{\phi}$.
As the rotation matrix $R \in SO(3)$ is governed by the condition $R R^T = R^TR = I$, we can take the derivative of this condition w.r.t. a parameter value $\phi_j$:
\begin{equation}
\begin{split}
\frac{\partial }{ \partial \phi_j} (RR^T) &= 0 , \\
\frac{\partial R }{ \partial \phi_j} R^T + R (\frac{\partial R }{ \partial \phi_j})^T &= 0,\\
\frac{\partial R }{ \partial \phi_j} R^T = - R (\frac{\partial R }{ \partial \phi_j})^T &= - [ \frac{\partial R }{ \partial \phi_j} R^T ]^T \in so(3),\\
\end{split}
\label{eq:RRT}
\end{equation}
where $ (\partial R / \partial \phi_j) R^T $ is a skew-symmetric matrix. Let 
\begin{equation}
\begin{split}
\hat{w}_{\phi_j} = (\partial R / \partial \phi_j) R^T \in so(3),\\ 
\end{split}
\label{eq:w_phi}
\end{equation}
where\ $\hat{} $ denotes the skew-symmetric operator that maps a vector in $\mathbb{R}^3$ to $so(3)$, we can then compute the rotational derivative w.r.t. the parameter $\phi_j$ as
\begin{equation}
\begin{split}
\frac{\partial R }{ \partial \phi_j} = \hat{w}_{\phi_j} R \in \mathbb{R}^{3\times 3}.\\ 
\end{split}
\label{eq:R_phi}
\end{equation}

Let ${w}_{\phi} \in \mathbb{R}^{3n_{\phi}}$ denote the collection of the virtual angular velocities for all parameters in $\phi$, we can then integrate ${w}_{\phi}$ along the catheter body for the calculation of rotational derivatives w.r.t. the parameters, where initial virtual angular velocity ${w}_{\phi,0} = \mathbf{0}$ at arc length $s_0$.
% and the initial value parameter vector is augmented as $\phi_2 = [p_0,\ u_0,\ {w}_{\phi}_0]^T $. 

\subsection{Flexible Segment IVP Parameter Derivatives}
% Let $X=[p,  u, R']^T$ denotes the concatenated vector of the position, curvature and rotation $R'$ at a given arc length $s$, where 
The CRM differential equations of a flexible segment can be expressed as
 \begin{equation}
\begin{split}
\frac{\partial X}{ \partial s} = F(X(s, \phi),\ s,\ \phi), % the CRM differential eqs 
\end{split}
\label{eq:CRM_differential}
\end{equation}
We can take the derivative of (\ref{eq:CRM_differential}) w.r.t. the parameters $\phi$:
\begin{equation}
\begin{split}
\frac{\partial }{ \partial \phi}\frac{\partial X}{ \partial s} = \frac{\partial F}{ \partial X} \frac{\partial X}{\partial \phi} + \frac{\partial F}{ \partial \phi},
\end{split}
\label{eq:X_s_phi}
\end{equation}
since $X$ is twice differentiable (as given in (\ref{eq:p_R_u_dot})), the derivative w.r.t. $s$ and $\phi$ is exchangeable:
\begin{equation}
\begin{split}
\frac{\partial }{ \partial s}\frac{\partial X}{ \partial \phi} = \frac{\partial F}{ \partial X} \frac{\partial X}{\partial \phi} + \frac{\partial F}{ \partial \phi}.\\
% \frac{\partial J_X}{ \partial s} = \frac{\partial F}{ \partial X} J_X + \frac{\partial F}{ \partial \phi}.
\end{split}
\label{eq:X_s_phi}
\end{equation}
$\partial X/ \partial \phi$ can then be integrated along the flexible segment for the computation of $J_{\mathrm{IVP}}$. The initial derivatives at the beginning of the flexible segment $s_0$ are all zero matrices, except that the initial derivatives of $\partial p/ \partial p_0$ and $\partial u/ \partial u_0$ are given as 3 by 3 identity matrices.
%  \begin{equation}
% \begin{split}
% \frac{\partial p}{ \partial p_0} |_{s = 0} &= \mathbf{I}^{3 \times 3},\
% \frac{\partial u}{ \partial u_0} |_{s = 0} =\mathbf{I}^{3 \times 3}.\\
% \end{split}
% \label{eq:initial_diff}
% \end{equation}

% and Equation (\ref{eq:P_phi_s_simplified}) can be reorganized using (\ref{eq:F_p}) and (\ref{eq:R_phi}):
% \begin{equation}
% \begin{split}
% \frac{\partial }{ \partial s}\frac{p}{ \partial \phi_i} = \left[ 
% \begin{array}{c}   
% \frac{\partial R_{13}}{\partial \phi_i}\\
% \frac{\partial R_{23}}{\partial \phi_i}\\
% \frac{\partial R_{33}}{\partial \phi_i}\\
%   \end{array}
% \right] = \left[ 
% \begin{array}{c}   
% (\hat{w}_{\phi}_i R)_{13}\\
% (\hat{w}_{\phi}_i R)_{23}\\
% (\hat{w}_{\phi}_i R)_{33}\\
%   \end{array}
% \right].    
% \end{split}
% \label{eq:p_phi_final}
% \end{equation}

\subsubsection{Positional Parameter Derivative}
The positional component of the differential equation (\ref{eq:X_s_phi}) can be computed as:
 \begin{equation}
\begin{split}
\frac{\partial }{ \partial s}\frac{\partial p}{ \partial \phi} &= \frac{\partial F_p}{ \partial X} \frac{\partial X}{\partial \phi} + \frac{\partial F_p}{ \partial \phi} \\
&= \frac{\partial F_p}{ \partial p} \frac{\partial p}{\partial \phi} + \frac{\partial F_p}{ \partial R} \frac{\partial R}{\partial \phi}  + \frac{\partial F_p}{ \partial u} \frac{\partial u}{\partial \phi} + \frac{\partial F_p}{ \partial \phi}, \\
\end{split}
\label{eq:P_phi_s}
\end{equation}
\vspace{-0.02cm}
where $F_p = \partial p / \partial s = R e_3$,
%  \begin{equation}
% \begin{split}
% \frac{\partial F_p}{ \partial p} = \frac{\partial F_p}{ \partial u} = \mathbf{0}^{3 \times 3}, \ \ \frac{\partial F_p}{ \partial \phi} = \mathbf{0}^{3 \times n_{\phi}}. \\
% \end{split}
% \label{eq:F_p}
% \end{equation}
and $\partial F_p/ \partial p = \partial F_p / \partial u =  \mathbf{0}^{3 \times 3}$, $\partial F_p /  \partial \phi = \mathbf{0}^{3 \times n_{\phi}}$.
(\ref{eq:P_phi_s}) is then simplified as:
 \begin{equation}
\begin{split}
\frac{\partial }{ \partial s}\frac{\partial p}{ \partial \phi} &= \frac{\partial F_p}{ \partial R} \frac{\partial R}{\partial \phi}  = \sum_{j=1}^{n_{\phi}} \frac{\partial F_p}{ \partial R} \frac{\partial R}{\partial \phi_j}  \\
&= \sum_{j=1}^{n_{\phi}} \sum_{m=1}^3\sum_{n=1}^3 \frac{\partial F_p}{\partial R_{mn}}  \frac{\partial R_{mn}}{\partial \phi_j}\\
% &= \sum_{j=1}^{n_X}  \left[ 
%      \frac{\partial R_{13}}{\partial \phi_j }\ \  \frac{\partial R_{23}}{\partial \phi_j }\ \  \frac{\partial R_{33}}{\partial \phi_j }
% \right]^T\\
&= \sum_{j=1}^{n_{\phi}}  \left[ \begin{array}{ccc}
     (\hat{w}_{\phi_j} R)_{13} & (\hat{w}_{\phi_j} R)_{23} & (\hat{w}_{\phi_j} R)_{33}
\end{array} \right]^T,
\end{split}
\label{eq:P_phi_s_simplified}
\end{equation}
% \vspace{-0.01cm}
where $(\hat{w}_{\phi_j} R)_{mn}$ denotes the element of $\hat{w}_{\phi_j} R$ at $m$-th row and $n$-th column.
\subsubsection{Curvature Parameter Derivative}
The curvature component of the differential equation (\ref{eq:X_s_phi}) is given as:
 \begin{equation}
\begin{split}
\frac{\partial }{ \partial s}\frac{\partial u}{ \partial \phi} 
&= \frac{\partial F_u}{ \partial p} \frac{\partial p}{\partial \phi} + \frac{\partial F_u}{ \partial R} \frac{\partial R}{\partial \phi}  + \frac{\partial F_u}{ \partial u} \frac{\partial u}{\partial \phi} + \frac{\partial F_u}{ \partial \phi}, \\
\end{split}
\label{eq:Fu_phi_s}
\end{equation}
where $F_u = \partial u / \partial s = u_s^* - K^{-1}( ( \hat{u}K + \dot{K} )(u-u^*) + \hat{e}_3R^T f_{cum} + R^T l )$.  The partial derivative of $F_u$ w.r.t. position is given as $\partial F_u/ \partial p = \mathbf{0}^{3 \times 3}$, and the partial derivative $F_u$ w.r.t. the curvature $u$ is computed as 
 \begin{equation}
\begin{split}
\frac{\partial F_u}{ \partial u} &= \frac{\partial }{ \partial u} [-K^{-1} \hat{u} K (u - u^*)] \\
&= K^{-1}[(Ku)^\wedge - (Ku^*)^\wedge - \hat{u}K]. \\
\end{split}
\label{eq:F_u_pRu}
\end{equation}
The partial derivative $\partial F_u / \partial \phi$ w.r.t. parameter vector $\phi$ is given as
 \begin{equation}
\begin{split}
\frac{\partial F_u}{ \partial f_{tip}} = -K^{-1}R^T, \ \
% \frac{\partial F_u}{ \partial z} = \mathbf{0}^{3 \times M}, \ 
\frac{\partial F_u}{ \partial u_0} &= K^{-1} ( \hat{u}K + \dot{K} ), 
% \frac{\partial F_u}{ \partial p_0} &= \mathbf{0}^{3 \times 3},\ \
% \frac{\partial F_u}{ \partial w_{\phi}_0} = \mathbf{0}^{3 \times 3n_{\phi}}\\
\end{split}
\label{eq:F_u_phi}
\end{equation}
and the initial derivatives of $F_u$ w.r.t. initial position $p_0$, initial virtual angular velocity $\partial w_{\phi,0}$, and actuation inputs $z$ are zero matrices.

We can calculate the derivatives of $F_u$ w.r.t. rotation vector using the virtual angular velocity $w_{\phi}$ as 
 \begin{equation}
\begin{split}
\frac{\partial F_u}{ \partial R} \frac{\partial R}{ \partial \phi} &= \sum_{j=1}^{n_{\phi}} \frac{\partial F_u}{ \partial R} \frac{\partial R}{ \partial \phi_j} \\ 
&= - \sum_{j=1}^{n_{\phi}} \sum_{m=1}^3\sum_{n=1}^3 K^{-1}(\hat{e}_3 \frac{\partial R^T}{\partial R_{mn}} f_{cum} + \frac{\partial R^T}{\partial R_{mn}} l ) \mathrlap{\frac{\partial R_{mn}}{\partial \phi_j}} \\
&= - \sum_{j=1}^{n_{\phi}} K^{-1}(\hat{e}_3 \frac{\partial R^T}{\partial \phi_j} f_{cum} + \frac{\partial R^T}{\partial \phi_j} l )\\
&= - \sum_{j=1}^{n_{\phi}} K^{-1}(\hat{e}_3 (\hat{w}_{\phi_j} R )^T f_{cum} + (\hat{w}_{\phi_j} R )^T l ).\\
\end{split}
\label{eq:F_u_R_PHI}
\end{equation}
\vspace{-0.2cm}

The curvature component of the parameter derivative can be computed by substituting (\ref{eq:F_u_pRu})-(\ref{eq:F_u_R_PHI}) into (\ref{eq:Fu_phi_s}).

\subsubsection{Rotational Parameter Derivative}

We can take the derivative of (\ref{eq:R_phi}) w.r.t. $s$ as 
\begin{equation}
\begin{split}
\frac{\partial }{ \partial s}\frac{\partial R }{ \partial \phi_j} &= (\frac{\partial }{\partial s} w_{\phi_j})^\wedge R +  \hat{w}_{\phi_j} \frac{\partial R}{\partial s},\\
% \frac{\partial }{ \partial s}\frac{\partial R }{ \partial \phi_j} &= (\frac{\partial }{\partial s} w_{\phi_j})^\wedge R +  \hat{w}_{\phi_j} R\hat{u}\\
(\frac{\partial }{\partial s} w_{\phi_j})^\wedge R &= \frac{\partial }{ \partial s}\frac{\partial R }{ \partial \phi_j}  - \hat{w}_{\phi_j} R\hat{u},\\
\frac{\partial }{\partial s} w_{\phi_j} &= [(\frac{\partial }{ \partial s}\frac{\partial R }{ \partial \phi_j}) R^T - \hat{w}_{\phi_j} R\hat{u} R^T]^\vee , \\
\end{split}
\label{eq:w_s}
\end{equation}
where the operator $^\vee$ maps a matrix from so(3) to a vector in $\mathbb{R}^3$. The first term on the right hand side can be expanded using the rotation matrix $R$ as 
 \begin{equation}
\begin{split}
\frac{\partial }{ \partial s}\frac{\partial R}{ \partial \phi_j} 
&= \frac{\partial F_{R}}{ \partial p} \frac{\partial p}{\partial \phi_j} + \frac{\partial F_{R}}{ \partial R} \frac{\partial R}{\partial \phi_j}  + \frac{\partial F_{R}}{ \partial u} \frac{\partial u}{\partial \phi_j} + \frac{\partial F_{R}}{ \partial \phi_j}, \\
\end{split}
\label{eq:R_phi_s}
\end{equation}
where $F_R = \partial R/  \partial s = R \hat{u}$. The partial derivatives $\partial F_{R} / \partial p$ and $\partial F_{R} / \partial \phi_j$ are zero matrices. The second and the third term on the right hand side can be computed as 
 \begin{equation}
\begin{split}
% F_R : \frac{\partial R}{ \partial s} &= R \hat{u}. 
% \frac{\partial F_{R}}{ \partial p} = \mathbf{0}^{ 3 \times 3 \times 3 },\ \frac{\partial F_{R}}{ \partial \phi_j} = \mathbf{0}^{3 \times 3  \times n_{\phi_j}}. \\
\frac{\partial F_R}{ \partial R} \frac{\partial R}{ \partial \phi_j} &=
\sum_{m=1}^3\sum_{n=1}^3 \frac{\partial F_R}{ \partial R_{mn}} \frac{\partial R_{mn}}{ \partial \phi_j}\\
 &= \sum_{m=1}^3\sum_{n=1}^3 (\frac{\partial R}{ \partial R_{mn}} \hat{u}) \frac{\partial R_{mn}}{ \partial \phi_i} = \hat{w}_{\phi_j}R \hat{u},\\
 \frac{\partial F_R}{ \partial u} \frac{\partial u}{ \partial \phi_j} &=
\sum_{m=1}^3 R( \frac{\partial u}{ \partial u_{m}} )^\wedge \frac{\partial u_{m}}{ \partial \phi_j} = R( \frac{\partial u}{ \partial \phi_j } )^\wedge. \\
\end{split}
\label{eq:R_pru_phi}
\end{equation}
(\ref{eq:R_phi_s}) can then be simplified as
 \begin{equation}
\begin{split}
\frac{\partial }{ \partial s}\frac{R}{ \partial \phi_j} &= \frac{\partial F_R}{ \partial R} \frac{\partial R}{ \partial \phi_j}  + \frac{\partial F_R}{ \partial u} \frac{\partial u}{ \partial \phi_j}\\
&= \hat{w}_{\phi_j}R \hat{u} + R( \frac{\partial u}{ \partial \phi_j } )^\wedge.
\end{split}
\label{eq:FR_R_Phi}
\end{equation}

We can substitute (\ref{eq:FR_R_Phi}) into (\ref{eq:w_s}):
 \begin{equation}
\begin{split}
\frac{\partial w_{\phi_j}}{ \partial s} &= [ (\frac{\partial }{ \partial s}\frac{\partial R }{ \partial \phi_j}) R^T - \hat{w}_{\phi_j} R\hat{u} R^T]^\vee\\
&= [ \hat{w}_{\phi_j}R \hat{u} R^T  + R( \frac{\partial u}{ \partial \phi_j } )^\wedge R^T - \hat{w}_{\phi_j} R\hat{u} R^T]^\vee\\
&= [ R( \frac{\partial u}{ \partial \phi_j } )^\wedge R^T]^\vee 
=[(R \frac{\partial u}{ \partial \phi_j })^\wedge ]^\vee = R \frac{\partial u}{ \partial \phi_j }
\end{split}
\label{eq:F_R_R_Phi}
\end{equation}

\subsection{IVP Parameter Derivatives Passed over Rigid Segments}

% The catheter state variable $X$ is passed through the rigid link using the following equations:
%  \begin{equation}
% \begin{split}
% p_{s_0, i+1} &= p_{s_f, i-1} + l_r *Re_3, \ \  
% R_{s_0, i+1} = R_{s_f, i-1}, \\
% u_{s_0, i+1} &= u_{s_0, i+1}^0 + K_{i+1}^{-1} K_{i-1}(u_{s_f, i-1} - u_{s_f, i-1}^0 ) - K_{i+1}^{-1} \tau_i\\ 
% \end{split}
% \label{eq:rigid_PRu}
% \end{equation}

Let $R_i$ denote the rotation matrix of the $i$-th rigid segment, we can take the derivatives of (\ref{eq:rigid_PRu}) w.r.t. the parameter values $\phi_j$:
 \begin{equation}
\begin{split}
\frac{\partial p_{s_0, i+1} }{\partial \phi_j} &= \frac{\partial p_{s_f, i-1} }{\partial \phi_j}+ L_i  \frac{\partial }{\partial \phi_j} R_ie_3\\
&= \frac{\partial p_{s_f, i-1} }{\partial \phi_j}+ L_i  [\hat{w}_{\phi_j} R_i]_{\mathbf{3}}, \\
\end{split}
\label{eq:rigid_PRu_phi}
\end{equation}
where $[\hat{w}_{\phi_j} R_i]_{\mathbf{3}}$ denotes the third column of $\hat{w}_{\phi_j} R_i$. Since the orientation remains the same at the start and the end of the rigid link, the rotational derivative is preserved as 
 \begin{equation}
\begin{split}
\frac{\partial R_{s_0, i+1} }{\partial \phi_j} &= \frac{\partial R_{s_f, i-1} }{\partial \phi_j} = \frac{\partial R_i}{\partial \phi_j} , \\
\end{split}
\label{eq:rigid_w_s}
\end{equation}
the virtual angular velocity vector at the end of the rigid link is then given as
 \begin{equation}
\begin{split}
% {w}_{\phi_j}_{s_0, i+1} &= {w}_{\phi_j}_{s_f, i-1}.
{w}_{\phi_j,\,s_0,i+1} = {w}_{\phi_j,\,s_f,i-1}.
\end{split}
\label{eq:rigid_w_s}
\end{equation}
Finally, we can compute the curvature derivative at the end of the given rigid link by taking derivative of curvature component of (\ref{eq:rigid_PRu}) as
 \begin{equation}
\begin{split}
\frac{\partial u_{s_0, i+1} }{\partial \phi_j} &= K_{i+1}^{-1} K_{i-1} \frac{\partial u_{s_f, i-1}}{\partial \phi_j} - K_{i+1}^{-1} \frac{d \tau_i}{d \phi_j}. \\
\end{split}
\label{eq:rigid_u_phi}
\end{equation}
The derivative of the magnetic moment $\tau_i$ w.r.t. the parameter values can be computed as
 \begin{equation}
\begin{split}
\frac{d \tau_i(\phi, R_i)}{d \phi_j} &= \frac{\partial \tau_i}{\partial \phi_j} + \frac{\partial \tau_i}{\partial R_i} \frac{\partial R_i}{\partial \phi_j}, \\
\end{split}
\label{eq:rigid_tau}
\end{equation}
where the partial derivative of $\tau_i$ w.r.t. the parameters are only non zero for the actuation currents
 \begin{equation}
\begin{split}
\frac{\partial \tau_i(\phi, R_i)}{\partial z^c} &=  -{B_i}^\wedge\, (N_i A_i).\\
% \frac{\partial \tau_i}{\partial p_0} &= \frac{\partial \tau_i}{\partial u_0} = \frac{\partial \tau_i}{\partial f_{tip}} = \mathbf{0}^{3 \times 3} \\
\end{split}
\label{eq:rigid_tau_i}
\end{equation}
 The second term of (\ref{eq:rigid_tau}) can be computed as
 \begin{equation}
\begin{split}
\frac{\partial \tau_i}{\partial R_i} \frac{\partial R_i}{\partial \phi_j} &= -(N_i A_i z_{i}^c)^\wedge R_i^T \hat{w}_{\phi_j} B_s.\\
\end{split}
\label{eq:rigid_tau_i}
\end{equation}
The curvature derivative can then be passed to the end of the rigid segment by substituting (\ref{eq:rigid_tau})-(\ref{eq:rigid_tau_i}) into (\ref{eq:rigid_u_phi}).

\subsection{Overall Forward Kinematic Jacobians}
Given IVP parameter derivatives $\partial X_{IVP} / \partial \phi$, the overall forward kinematic Jacobians $J_{\mathrm{BVP}}$ that relates the catheter tip configurations to the catheter control input $z$ can then be calculated by
\begin{equation}
J_{\mathrm{BVP}} = \frac{\partial X_{\mathrm{BVP}}}{\partial z} = \frac{\partial X_{IVP} }{\partial z}  + \frac{\partial X_{IVP} }{\partial u_0} \frac{d u_0}{ d z}. 
\label{eq:p_mu_part1}
\end{equation}
where $X_{\mathrm{BVP}} = [p_{\mathrm{BVP}},\ R_{\mathrm{BVP}},\ u_{\mathrm{BVP}}]$ denotes the catheter states solved by the BVP function in (\ref{eq:BVP_function}).

%% file: experiment.tex
\section{Experimental Validation}\label{section:experiment}

% % \input{result} 
In this section, the performance of the proposed kinematic model and the Jacobian calculation are experimentally vali- dated using a single coil set catheter prototype. The proposed forward kinematics and Jacobian calculations are implemented using the iterative-Jacobian based open-loop control method. The accuracy, repeatability, and computational efficiency of the proposed approach are evaluated in experiments.

\subsection{Iterative Jacobian Based Open Loop Control}\label{sec:iterativeJacobianInverseKinematics}

In this paper, the robotic catheter is controlled in open-loop using the iterative Jacobian based method \cite{liu2017iterative}, where the forward kinematics Jacobian is employed to iteratively update the control inputs. Let $J_{\mathrm{BVP}, p}$ denote the positional BVP Jacobian\footnote{The readers are referred to \cite{itsarachaiyot2023analytical} for the complete derivation of the BVP Jacobians}, the changes in the catheter tip position can then be related to the changes in the actuation currents as:
\begin{equation}
dp = J_{\mathrm{BVP}, p} dz.
\label{eq:posiitoncurrents}
\end{equation}
When the changes in the catheter tip position are small, the desired change in actuation currents to achieve a given tip displacement using the inverse of the BVP Jacobian, using a linear approximation as:
\begin{equation}
dz  =J_{\mathrm{BVP}}^T ( J_{\mathrm{BVP}} J_{\mathrm{BVP}}^T +\lambda^2 I_{3x3} )^{-1} dp,
\label{eq:inversekinematics}
\end{equation}
where the damped least-square method is used to avoid the instability issues \cite{liu2017iterative,more2006levenberg,buss2004introduction} and $\lambda$ is a nonzero damping constant.

\subsection{Accuracy and Reproducibility Evaluation}\label{subsection:model_performance}

\begin{figure*}
\centering
\begin{subfigure}[t]{.24\textwidth}
  \centering
  \includegraphics[width=\textwidth]{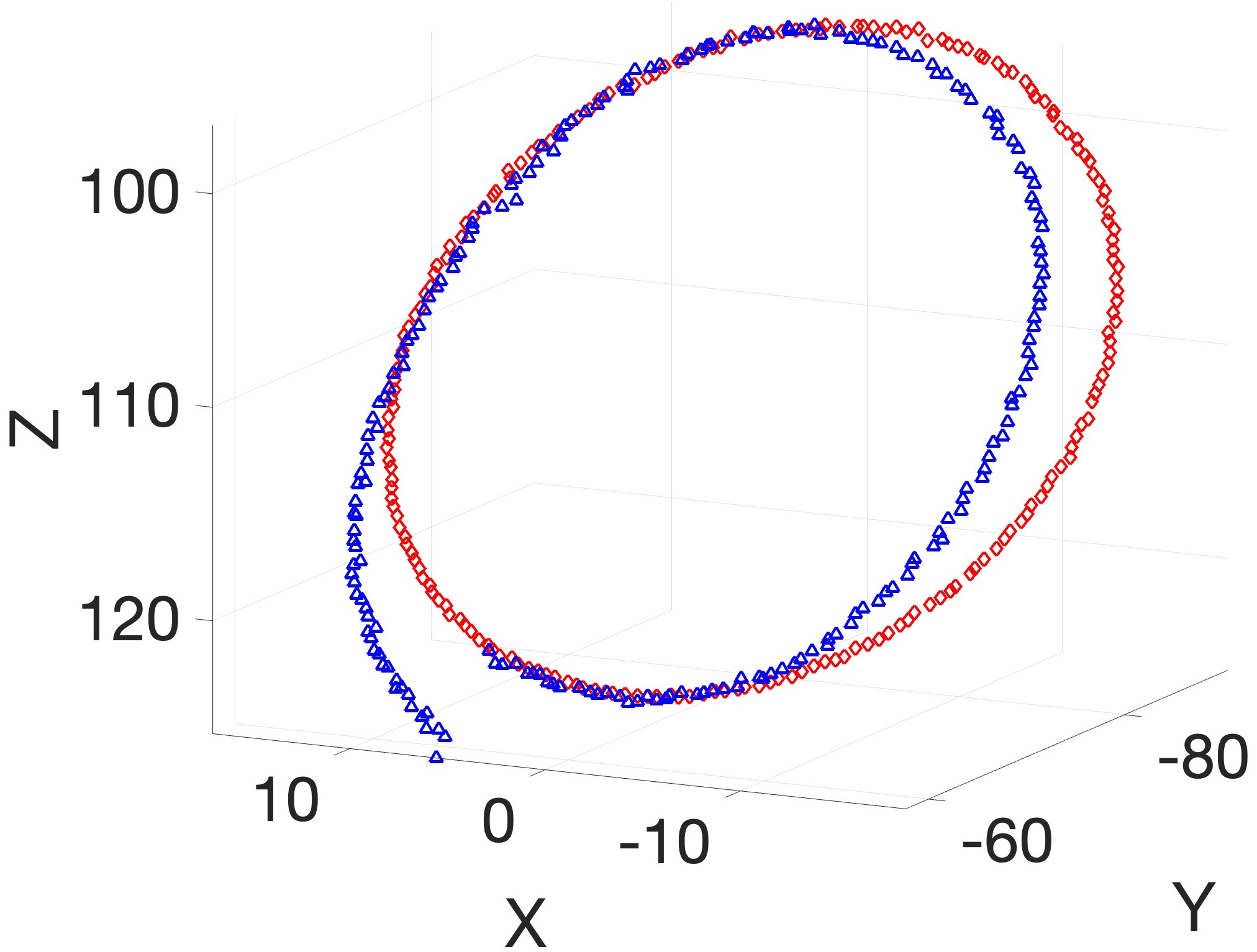}
  \caption{}
  \label{fig:sfig1}
\end{subfigure}
% \hfill
\begin{subfigure}[t]{.24\textwidth}
  \centering
  \includegraphics[width=\textwidth]{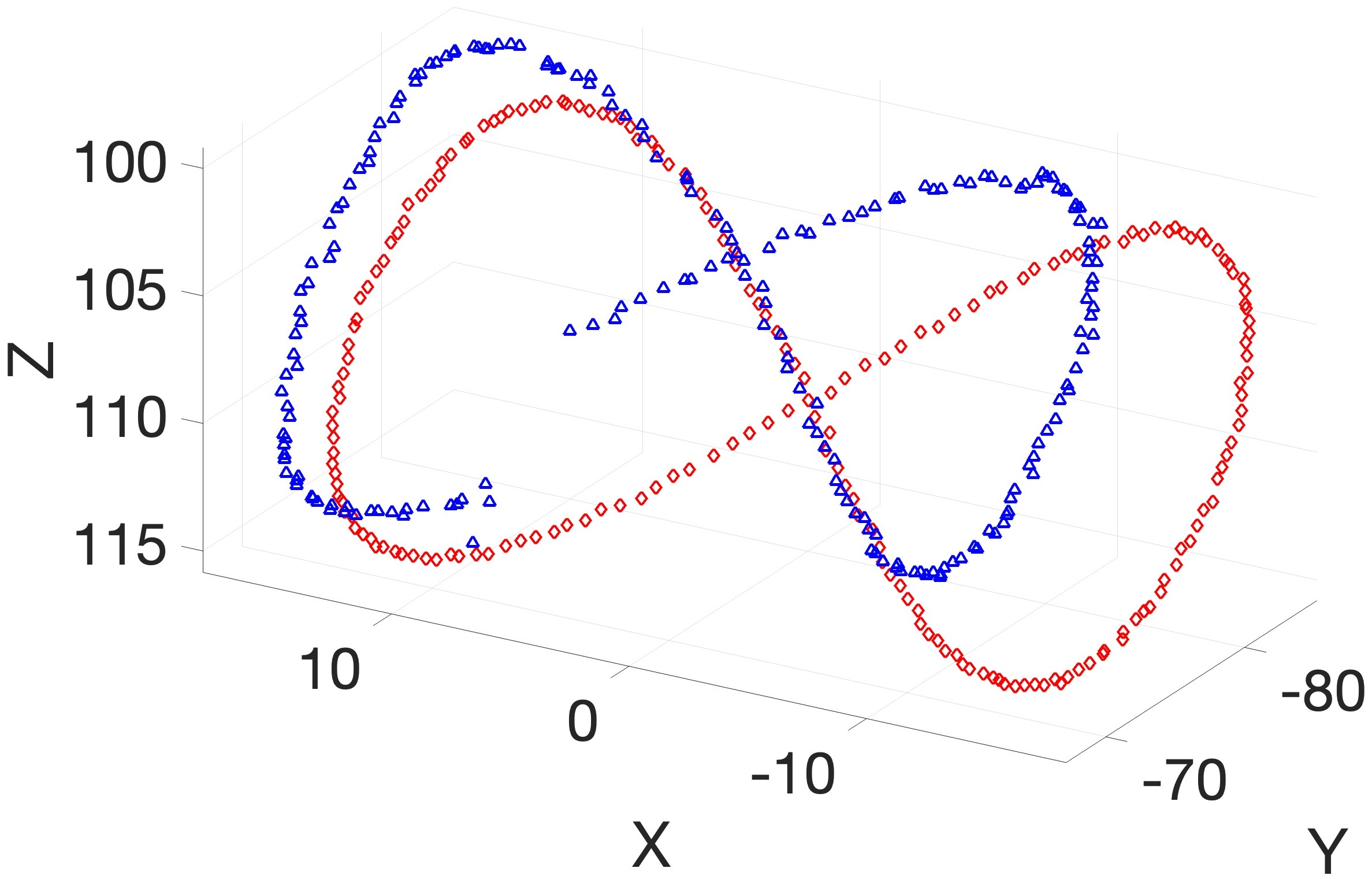}
  \caption{}
  \label{fig:sfig2}
\end{subfigure}
% \hfill
\begin{subfigure}[t]{.24\textwidth}
  \centering
  \includegraphics[width=\textwidth]{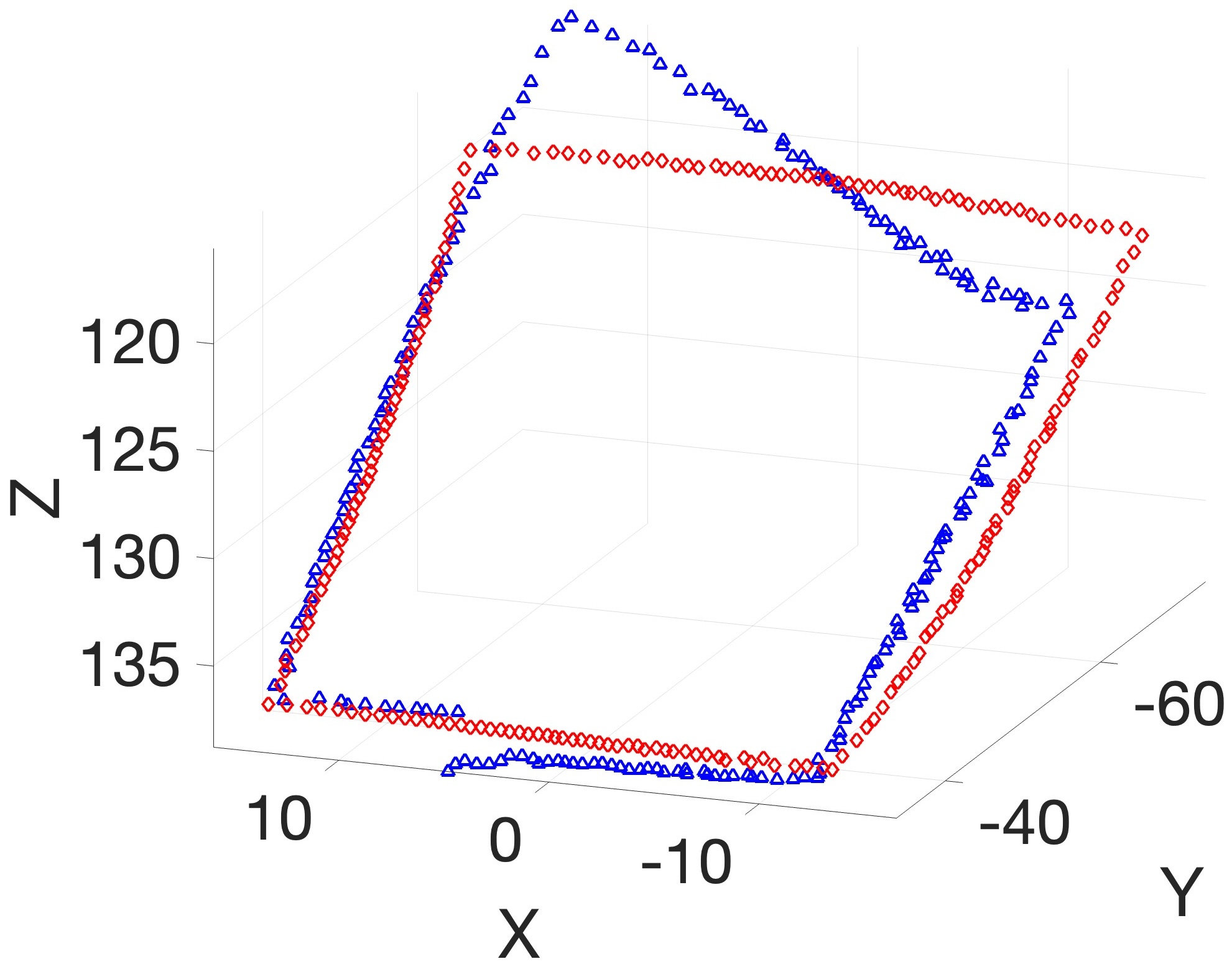}
  \caption{}
  \label{fig:sfig3}
\end{subfigure}
% \hfill
\begin{subfigure}[t]{.24\textwidth}
  \centering
  \includegraphics[width=\textwidth]{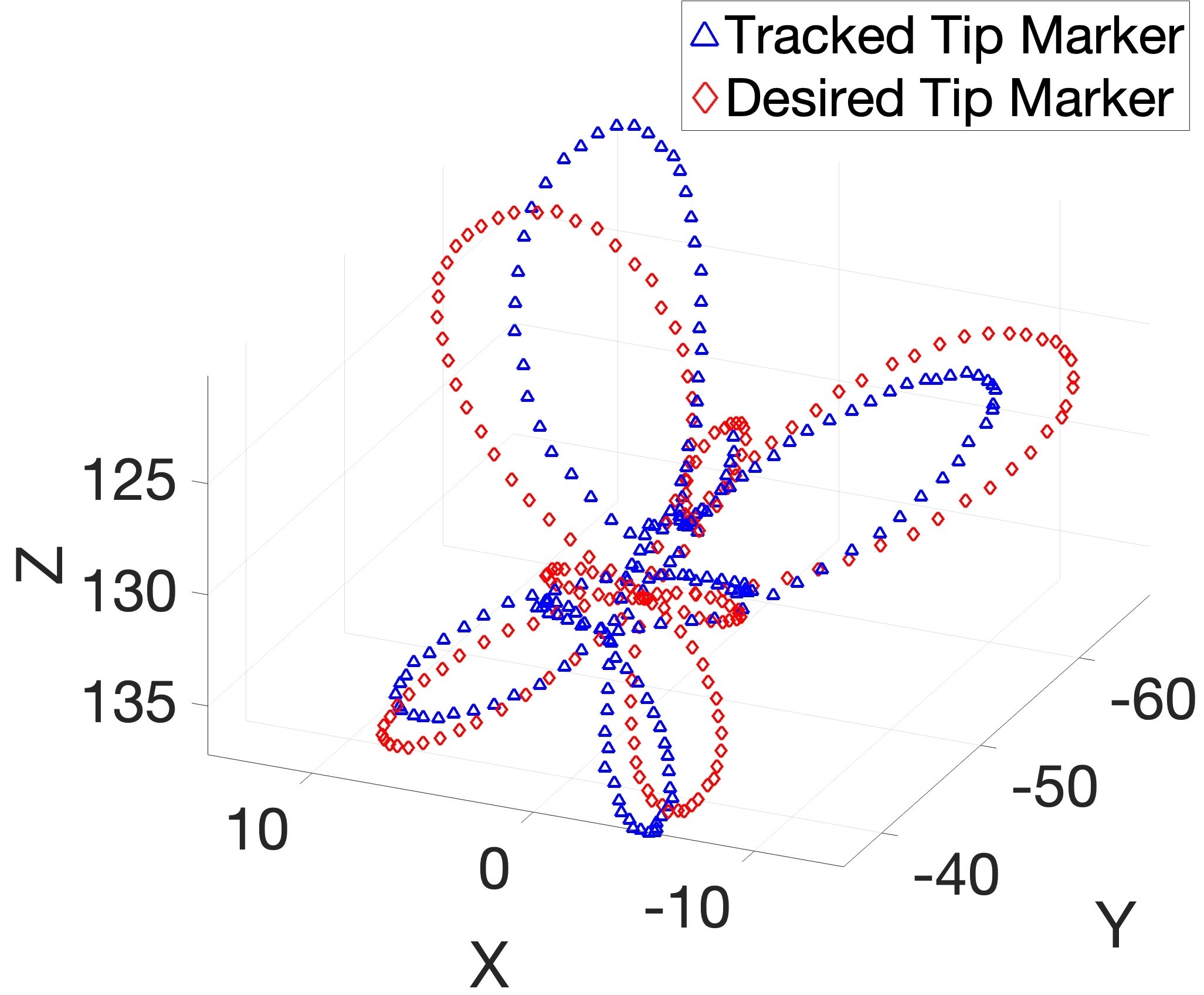}
  \caption{}
  \label{fig:sfig4}
\end{subfigure}
\caption{Open-loop trajectories generated using the Cosserat-rod model. The desired tip trajectories are represented by red diamonds. The observed trajectories of the tip location collected using the catadioptric stereo tracking system are represented by blue triangles. All units are in mm. (a) Circle trajectory. (b) Lemniscate trajectory. (c) Rectangle trajectory. (d) Butterﬂy trajectory.}
\label{fig:example_trajectory}
\end{figure*}

In this section, the proposed kinematic model and Jacobian calculations are experimentally validated using 4 distinct sets of control input trajectories, i.e., a circle trajectory, a lemniscate trajectory, a rectangle trajectory, and a butterfly trajectory. The robotic catheter is controlled using the iterative-Jacobian based method in an open-loop fashion, where the catheter tip trajectories are recorded and tracked by the catadioptric stereo tracking system. Fig.~\ref{fig:example_trajectory} shows the 4 sets of catheter tip trajectories under the 4 given sets of open-loop control inputs. The desired catheter tip positions are represented using the red diamonds and the observed trajectories of the tip positions collected using the catadioptric stereo tracking system are represented by the blue triangles. The tip positional difference is used as a metric for the accuracy evaluation of the trajectories generated using the iterative-Jacobian based control approach. The root-mean-square errors (RMSEs) between the open-loop trajectories generated using the proposed method and the actual trajectories collected using the catadioptric stereo tracking system, are used as the quantitative metric for performance evaluation.

To evaluate the accuracy of the open-loop control, the catheter tip trajectories generated using the proposed kinematic model are evaluated against the observed trajectories of the tip location collected using the catadioptric stereo tracking system, as shown in Fig.~\ref{fig:example_trajectory}. As the catheter is controlled in an open-loop fashion, where no positional feedback measurement is used during the control, errors between the desired and the observed trajectories are expected. The RMSEs between the desired and observed trajectories are reported in Table~\ref{tb:table_accuracy_crm_pebax}. 

To evaluate the repeatability of the proposed modeling approach, each trajectory is repeated 10 times where the catheter tip positions are collected using the catadioptric stereo tracking system. The open-loop trajectory repeatability was quantified in two ways. First, the observed catheter tip trajectories are compared with the desired tip trajectories by ignoring the position and orientation offsets caused by the open-loop control errors. Specifically, the trajectories are translated and rotated (but not scaled) until the RMSEs among the trajectories are minimized and the residual RMSEs are used as the trajectory similarity measure. This is to ensure that the resulting repeatability error is not sensitive to offset (and some drift) types of errors that are inherent to open- loop control schemes, which can be easily eliminated with closed-loop control. Second, a reference trajectory randomly selected from the 10 observed trajectories is compared with the observed catheter tip trajectories for repeatability evaluation. These two sets of results are presented in Table~\ref{tb:table_reproduce_crm_pebax}.

\begin{table}[tb]
\caption{RMSEs between the generated trajectories using the Cosserat-rod based kinematic model and the observed trajectories. }
\label{tb:table_accuracy_crm_pebax}
\begin{center}
\def\arraystretch{1.30}
\begin{tabular}{l c c c c}
\hline\hline 
Trajectory & Circle & Lemniscate & Rectangle & Butterfly\\ [0.5ex]
\hline
Mean &  4.34 &  6.49 &  4.60 &  3.66\\
Variance &  0.0581 &  0.0063 &  0.0034 &  0.0668\\
% Trial 01 &  4.89 &  6.51 &  4.48 &  3.32\\
% Trial 02 &  4.10 &  6.53 &  4.54 &  3.33\\
% Trial 03 &  4.03 &  6.39 &  4.56 &  3.46\\
% Trial 04 &  4.46 &  6.38 &  4.60 &  3.62\\
% Trial 05 &  4.36 &  6.57 &  4.59 &  3.75\\
% Trial 06 &  4.07 &  6.52 &  4.63 &  3.47\\
% Trial 07 &  4.36 &  6.62 &  4.64 &  3.65\\
% Trial 08 &  4.41 &  6.39 &  4.66 &  3.88\\
% Trial 09 &  4.48 &  6.54 &  4.66 &  4.00\\
% Trial 10 &  4.22 &  6.44 &  4.67 &  4.10\\ [1.0ex]
\hline\hline
\end{tabular}
\end{center}
\footnotesize{The means and variances are calculated from the RMS errors. All units are in mm.}
\end{table}

\begin{table*}[tb]
\caption{In the first set of results, the shapes of the observed trajectories are compared to the given desired trajectory, calculated as the RMSEs between the observed and the desired tip trajectories, ignoring the position and orientation offsets. The second set of results reports the repeatability of the observed trajectories, calculated as the RMSEs between a randomly selected observed trajectory and all of the other observed trajectories.}
\label{tb:table_reproduce_crm_pebax}
\begin{center}
\def\arraystretch{1.30}
\begin{tabular}{l c c c c c c c c c}
\hline\hline 
& \multicolumn{4}{c}{Trajectory Accuracy Ignoring Position and Orientation Offsets} && \multicolumn{4}{c}{Trajectory Repeatability}\\
\cline{2-5}\cline{7-10}
Trajectory & Circle & Lemniscate & Rectangle & Butterfly && Circle & Lemniscate & Rectangle & Butterfly\\ [0.5ex]
\hline
Mean &  1.81 &  2.29 &  3.35 &  2.72 && 0.83 &  0.64 &  0.44 &  0.59\\
Variance &  0.0064 &  0.0130 &  0.0017 &  0.0013 && 0.0792 &  0.0059 &  0.0003 &  0.0102\\
% Trial 01 &  1.98 &  2.56 &  3.24 &  2.82 && - &  - &  - &  -\\
% Trial 02 &  1.83 &  2.40 &  3.32 &  2.69 && 0.34 &  0.53 &  0.40 &  0.46\\
% Trial 03 &  1.81 &  2.33 &  3.35 &  2.72 && 0.42 &  0.52 &  0.44 &  0.48\\
% Trial 04 &  1.77 &  2.31 &  3.36 &  2.70 && 0.86 &  0.58 &  0.46 &  0.60\\
% Trial 05 &  1.69 &  2.25 &  3.34 &  2.71 && 0.78 &  0.66 &  0.44 &  0.63\\
% Trial 06 &  1.69 &  2.24 &  3.37 &  2.74 && 0.73 &  0.65 &  0.46 &  0.44\\
% Trial 07 &  1.82 &  2.24 &  3.36 &  2.71 && 0.99 &  0.72 &  0.43 &  0.65\\
% Trial 08 &  1.88 &  2.20 &  3.38 &  2.72 && 1.09 &  0.65 &  0.45 &  0.68\\
% Trial 09 &  1.86 &  2.16 &  3.38 &  2.71 && 1.20 &  0.76 &  0.45 &  0.69\\
% Trial 10 &  1.79 &  2.18 &  3.39 &  2.70 && 1.09 &  0.69 &  0.42 &  0.71\\ [1.0ex]
\hline\hline
\end{tabular}
\end{center}
\footnotesize{All units are in millimeter.}
\end{table*}

\begin{figure}
  \centering
  \includegraphics[scale=0.22]{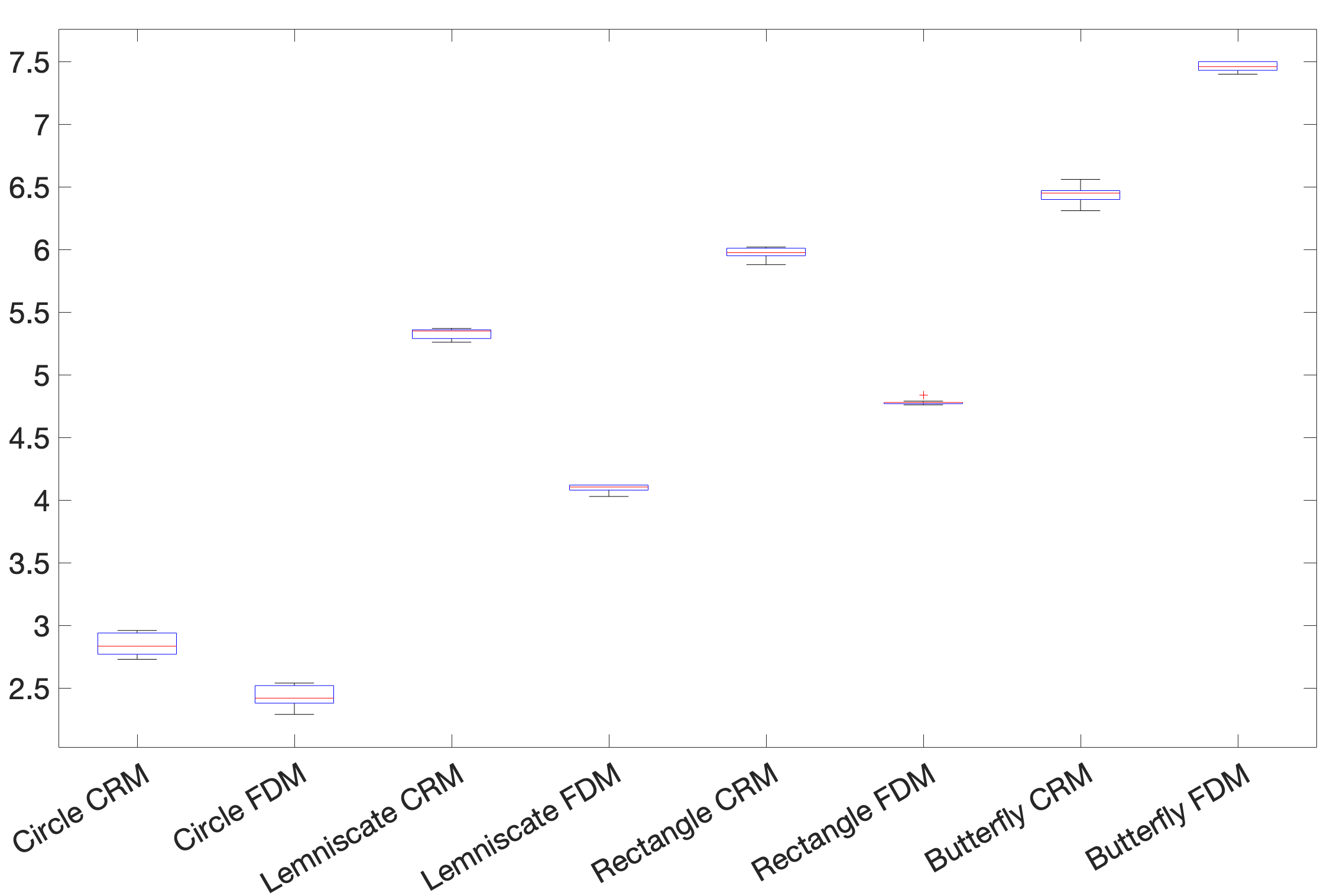}
  \caption{The mean and variance of the RMSEs between the observed trajectories and desired trajectories generated using the Cosserat-rod based kinematic model (CRM) and the finite difference based kinematic model (FDM) over 10 trials for circle, lemniscate, rectangle, and butterfly trajectories.}
  \label{pic:comparison}
\end{figure}

The accuracy and reproducibility of the open-loop control method implemented based on the Cosserat-rod based kinematic model (CRM) is evaluated against the same open-loop control method implemented based on the finite-difference based kinematic model (FDM) proposed in \cite{liu2017iterative}. The mean and variance of the RMSEs between 10 observed trajectory shapes and the desired trajectory shape generated using the Cosserat-rod based kinematic model and the finite-difference based kinematic model for circle, lemniscate, rectangle, and butterfly trajectories, are presented in Fig.~\ref{pic:comparison}.  

% The results of the RMSEs of both reproducibility and accuracy performance indicate that the overall shape of the experimental observed trajectory has very good similarity to the corresponding desired trajectory. 

% \subsection{Kinematic Model Performance Comparison}\label{subsection:model_performance_comparison}

\subsection{Computational Efficiency of the Analytical Jacobian Derivation}

The proposed analytical Jacobian derivation and kinematic model are implemented in C++ on Ubuntu 20.04 operating system. The computer is equipped with Intel\textsuperscript{\textregistered} Core\textsuperscript{TM} i9-11900 CPU @ 2.50GHz and 32.0 GiB memory. 
The average computational time for the virtual angular velocity based Jacobian derivation $J_{\mathrm{BVP}}$ and finite-element based Jacobian derivation \cite{rucker2011computing} are compared for 100 randomly sampled catheter configurations under different actuation inputs.
% \begin{figure}[t]
% \includegraphics[scale=0.45]{pictures/100samples_0.3A.jpg}
% \centering
% \caption{100 randomly sampled catheter configurations under different control inputs.}
% \label{pic:100samples}
% \end{figure}
In addition, the computational efficiency of the proposed forward kinematic model implemented using the virtual angular velocity based Jacobian calculation is evaluated against the finite-element based forward kinematic model \cite{liu2016modeling}. The average run time per step along the 4 trajectory sets shown in Fig.~\ref{fig:example_trajectory} of applying the iterative-Jacobian-based control implemented with Cosserat-rod-based kinematic model and the finite-difference-based kinematic model proposed in \cite{liu2017iterative} are compared.

% As the run time of the kinematics is crucial for clinical procedures. 
The computational efficiency for the proposed virtual angular velocity Jacobian derivation was evaluated under 100 samples and compared against the finite-element based derivation approach. The results show that the average computational time for calculating the forward kinematic Jacobians using the proposed analytical derivation is 2.0 ms, compared to the computational time of 5.5 ms using the numerical Jacobian calculation. This highlights the computational advantage of the virtual angular velocity method in real-time applications.
The average run time per step along the trajectory of applying the iterative-Jacobian-based open-loop control implemented using the proposed Jacobian calculation and kinematic model is 4.1 ms.
% For closed-loop control of the MRI-actuated robotic catheter with real-time MRI feedback information, the run time per step of the open-loop control should be significantly smaller than the targeted image acquisition time of 16 ms for catheter localization \cite{fransonPhD}.  
The proposed kinematic model and Jacobian calculation have a significantly faster run time, compared to the open-loop control implementation using the finite-difference-based kinematic model in \cite{liu2017iterative}, which has an average run time of 95 ms per step along the trajectory.